\documentclass[preprintnumbers,amsmath,amssymbm,prd]{revtex4}
\usepackage{epsfig}
\usepackage{graphicx}
\usepackage{amssymb}

\begin{document}
\title{Marginally stable resonant modes of the polytropic hydrodynamic vortex}
\author{Shahar Hod}
\affiliation{The Ruppin Academic Center, Emeq Hefer 40250, Israel}
\affiliation{ } \affiliation{The Hadassah Institute, Jerusalem
91010, Israel}
\date{\today}

\begin{abstract}
\ \ \ The polytropic hydrodynamic vortex describes an effective
$(2+1)$-dimensional acoustic spacetime with an inner reflecting
boundary at $r=r_{\text{c}}$. This physical system, like the
spinning Kerr black hole, possesses an ergoregion of radius
$r_{\text{e}}$ and an inner non-pointlike curvature singularity of
radius $r_{\text{s}}$. Interestingly, the fundamental ratio
$r_{\text{e}}/r_{\text{s}}$ which characterizes the effective
geometry is determined solely by the dimensionless polytropic index
$N_{\text{p}}$ of the circulating fluid. It has recently been proved
that, in the $N_{\text{p}}=0$ case, the effective acoustic spacetime
is characterized by an {\it infinite} countable set of reflecting
surface radii, $\{r_{\text{c}}(N_{\text{p}};n)\}^{n=\infty}_{n=1}$,
that can support static (marginally-stable) sound modes. In the
present paper we use {\it analytical} techniques in order to explore
the physical properties of the polytropic hydrodynamic vortex in the
$N_{\text{p}}>0$ regime. In particular, we prove that in this
physical regime, the effective acoustic spacetime is characterized
by a {\it finite} discrete set of reflecting surface radii,
$\{r_{\text{c}}(N_{\text{p}},m;n)\}^{n=N_{\text{max}}}_{n=1}$, that
can support the marginally-stable static sound modes (here $m$ is
the azimuthal harmonic index of the acoustic perturbation field).
Interestingly, it is proved analytically that the dimensionless
outermost supporting radius
${{r^{\text{max}}_{\text{c}}}/{r_{\text{e}}}}$, which marks the
onset of superradiant instabilities in the polytropic hydrodynamic
vortex, increases monotonically with increasing values of the
integer harmonic index $m$ and decreasing values of the
dimensionless polytropic index $N_{\text{p}}$.
\end{abstract}
\bigskip
\maketitle

%]

\section{Introduction}

One of the most intriguing phenomenon in general relativity is the
dragging of inertial frames by rotating bodies \cite{Chan}. This
fundamental physical effect is demonstrated most dramatically in the
spinning Kerr spacetime \cite{Kerr}. In particular, this black-hole
solution of the classical Einstein field equations is characterized
by an ergoregion \cite{Chan}, a spacetime region located between the
horizon and the static surface \cite{Notestat} in which physical
observers cannot appear static with respect to inertial observes at
asymptotic spatial infinity.

Interestingly, it has been demonstrated more than four decades ago
\cite{Zel,PressTeu1,PressTeu2} that energy and angular momentum can
be extracted from the ergoregion of a spinning black hole by a
co-rotating cloud of bosonic (integer-spin) fields. Nevertheless,
the Kerr spacetime solution of the Einstein field equations is known
to be stable to linearized perturbations of massless fields
\cite{PressTeu2,Whit,Cars}. This important physical feature of the
spinning black-hole spacetime is attributed to the fact that a
classical black-hole horizon acts as an {\it absorbing} one-way
membrane. Thus, massless perturbation fields are eventually
swallowed by the black hole or radiated away to infinity
\cite{PressTeu2,Whit,Cars,Notebom,Ins1,Ins2,Ins3}.

As opposed to the asymptotically flat Kerr black-hole spacetime,
whose stability to massless bosonic perturbations is guaranteed by
the {\it absorbing} nature of its horizon
\cite{PressTeu2,Whit,Cars}, horizonless spinning spacetimes with
{\it reflecting} boundaries and ergoregions may develop
characteristic superradiant instabilities to co-rotating bosonic
fields \cite{Fri,Cars,Carsup,Sla}. In particular, it has recently
been demonstrated \cite{Cars,Hods,CCP} that exponentially growing
superradiant instabilities may develop in the effective ergoregions
of rotating fluid systems \cite{Noteunr,Unr,Fed,Dol}.

The physically interesting polytropic hydrodynamic vortex studied in
\cite{CCP} describes an horizonless $(2+1)$-dimensional purely
circulating fluid with an effective acoustic ergoregion of radius
$r_{\text{e}}$ and an inner reflecting boundary of radius
$r_{\text{c}}$. As explicitly proved in \cite{Cars,Hods,CCP}, the
presence of an ergoregion together with the absence of an absorbing
horizon, guarantee that the effective acoustic spacetime described
by the rotating fluid is superradiantly unstable to linearized sound
perturbation modes.

Intriguingly, for a given value of the dimensionless polytropic
index $N_{\text{p}}$ which characterizes the circulating fluid [see
Eq. (\ref{Eq1}) below], the horizonless acoustic spacetime of the
hydrodynamic vortex is characterized by static ({\it
marginally-stable}) resonances that mark the boundary between stable
and superradiantly unstable spinning fluid configurations
\cite{Cars,Hods,CCP}. The main goal of the present paper is to study
{\it analytically} the physical and mathematical properties of these
characteristic marginally-stable acoustic spacetime resonances.

\section{Description of the system}

The polytropic hydrodynamic vortex describes a $(2+1)$-dimensional
purely circulating fluid which is characterized by the compact
pressure-density functional relation \cite{CCP}
\begin{equation}\label{Eq1}
P(\rho)=k_{\text{p}}\rho^{1+1/N_{\text{p}}}\  ,
\end{equation}
where the coefficient $k_{\text{p}}$ is the polytropic constant and
the dimensionless constant $N_{\text{p}}$ is the polytropic index of
the fluid \cite{Hor}. This physical system is described by the
effective acoustic geometry \cite{CCP}
\begin{equation}\label{Eq2}
ds^2={{\rho}\over{c_{\text{s}}}}\big[-c^2_{\text{s}}dt^2+(rd\theta-v_{\theta}dt)^2+dr^2+dz^2\big]\
,
\end{equation}
where the radially dependent physical parameters $\rho=\rho(r)$ and
$v_{\theta}=v_{\theta}(r)$ are respectively the mass density and the
angular velocity of the fluid, and $c_{\text{s}}$ is the propagation
speed of sound perturbation modes in the fluid.

Interestingly, as explicitly shown in \cite{CCP}, the effective
acoustic spacetime of the polytropic hydrodynamic vortex, like the
spinning Kerr black-hole spacetime, is characterized by a
non-pointlike singularity of radius $r_{\text{s}}$ \cite{Notepar}.
At this inner radius, the mass density $\rho$ of the fluid and the
speed $c_{\text{s}}$ of sound modes go to zero, signaling that the
scalar curvature of the corresponding acoustic geometry becomes
singular \cite{CCP}. In addition, the spinning acoustic spacetime
(\ref{Eq2}), which describes the polytropic hydrodynamic vortex, is
characterized, like the familiar Kerr black-hole spacetime, by an
ergoregion of radius $r_{\text{e}}$ \cite{CCP,Notecir}.

As explicitly proved in \cite{CCP}, the dimensionless ratio
$r_{\text{e}}/r_{\text{s}}$ between the radius of the ergoregion and
the radius of the inner singularity which characterizes the
polytropic hydrodynamic vortex is determined solely by the
polytropic index $N_{\text{p}}$ of the fluid \cite{CCP}:
\begin{equation}\label{Eq3}
{{r_{\text{e}}}\over{r_{\text{s}}}}=\sqrt{1+2N_{\text{p}}}\  .
\end{equation}
Below we shall focus on the physical regime $N_{\text{p}}>0$, in
which case one finds from (\ref{Eq3}) the inequality
$r_{\text{s}}<r_{\text{e}}$, a relation which also characterizes the
rotating Kerr black-hole spacetime.

The spatial and temporal properties of linearized perturbations
modes in the effective acoustic spacetimes (\ref{Eq2}) are governed
by the Klein-Gordon equation \cite{CCP,Unr,Cars,Notefl}
\begin{equation}\label{Eq4}
\nabla^\nu\nabla_{\nu}\Psi={{1}\over{\sqrt{|g|}}}\partial_{\mu}\Big(\sqrt{|g|}
g^{\mu\nu}\partial_{\nu}\Psi\Big)=0\  .
\end{equation}
We shall henceforth focus on the {\it static} perturbation modes of
the polytropic hydrodynamic vortex which, as nicely demonstrated
numerically in \cite{CCP}, mark the onset of rotational
instabilities in the acoustic curved spacetime (\ref{Eq2}).
Substituting the metric components of the curved line element
(\ref{Eq2}) into (\ref{Eq4}), defining the dimensionless radial
coordinate
\begin{equation}\label{Eq5}
x\equiv {{r}\over{r_{\text{s}}}}\  ,
\end{equation}
and using the mathematical decomposition \cite{Notemm}
\begin{equation}\label{Eq6}
\Psi(x,\phi,z)={{1}\over{\sqrt{x}}}\sum_{m=-\infty}^{\infty}\psi_m(x)e^{im\phi}\
,
\end{equation}
one obtains the radial differential equation \cite{CCP}
\begin{equation}\label{Eq7}
\Big[x^2(x^2-1){{d^2}\over{dx^2}}+2N_{\text{p}}x{{d}\over{dx}}
+(2m^2-1)N_{\text{p}}-(x^2-1)(m^2-{1\over4})\Big]\psi_m(x)=0\
\end{equation}
which determines the spatial behavior of the static
(marginally-stable) linearized perturbation modes that characterize
the polytropic hydrodynamic vortex. Note that the radial wave
equation (\ref{Eq7}) for the static resonant modes of the polytropic
hydrodynamic vortex is invariant under the reflection transformation
$m\to -m$. Hence, we shall henceforth assume
\begin{equation}\label{Eq8}
m>0\
\end{equation}
without loss of generality.

Interestingly, the ordinary differential equation (\ref{Eq7}) can be
solved analytically in terms of the familiar hypergeometric function
\cite{Abram}, yielding the physically acceptable radial
eigenfunction \cite{CCP,Abram,Notepa}
\begin{equation}\label{Eq9}
\psi_m(x;N_{\text{p}})=A\cdot x^{{{1}\over{2}}-m}{_2F_1}
\Big[{{N_{\text{p}}+m-\sqrt{N^2_{\text{p}}+m^2(1+2N_{\text{p}})}}\over{2}},
{{N_{\text{p}}+m+\sqrt{N^2_{\text{p}}+m^2(1+2N_{\text{p}})}}\over{2}};
1+m;x^{-2}\Big]\
\end{equation}
for the static (marginally-stable) resonances of the polytropic
hydrodynamic vortex, where ${_2F_1}(a,b;c;z)$ is the hypergeometric
function \cite{Abram} and $A$ is a normalization constant.

In order to exclude the curvature singularity from the acoustic
spacetime, Oliveira, Cardoso, and Crispino \cite{CCP} have suggested
to place an infinitely long {\it reflecting} cylinder of radius
$r_{\text{c}}$, which is characterized by the inequality
\begin{equation}\label{Eq10}
r_{\text{c}}\geq r_{\text{s}}\  ,
\end{equation}
at the center of the cylindrically symmetric [an effective
$(2+1)$-dimensional] physical system. In particular, in order to
allow a fully {\it analytical} treatment of the polytropic
hydrodynamic vortex, we shall assume that the characteristic radial
eigenfunction (\ref{Eq9}) of the linearized acoustic perturbation
modes vanishes on the surface $r=r_{\text{c}}$ of the central
reflecting cylinder \cite{CCP}:
\begin{equation}\label{Eq11}
\psi(r=r_{\text{c}})=0\  .
\end{equation}
Substituting (\ref{Eq9}) into (\ref{Eq11}) and defining
\begin{equation}\label{Eq12}
x_{\text{c}}\equiv {{r_{\text{c}}}\over{r_{\text{s}}}}\  ,
\end{equation}
one obtains the characteristic resonance condition
\begin{equation}\label{Eq13}
{_2F_1}\Big[{{N_{\text{p}}+m-\sqrt{N^2_{\text{p}}+m^2(1+2N_{\text{p}})}}\over{2}},
{{N_{\text{p}}+m+\sqrt{N^2_{\text{p}}+m^2(1+2N_{\text{p}})}}\over{2}};
1+m;x^{-2}_{\text{c}}\Big]=0\  ,
\end{equation}
which determines the discrete set
$\{x_{\text{c}}(N_{\text{p}},m;n)\}$ of dimensionless cylinder radii
that can support the marginally-stable (static) acoustic
perturbation modes of the polytropic hydrodynamic vortex.

Previous explorations \cite{Cars,Hods} of the effective
$(2+1)$-dimensional circulating fluid system have revealed the
physically interesting fact that, in the $N_{\text{p}}=0$ case, the
acoustic spacetime is characterized by an {\it infinite} countable
set of reflecting cylinder radii,
$\{r_{\text{c}}(N_{\text{p}}=0,m;n)\}^{n=\infty}_{n=1}$, that can
support the static (marginally-stable) resonant acoustic modes. In
the present paper we shall use {\it analytical} techniques in order
to study the characteristic resonance condition (\ref{Eq13}) of the
polytropic hydrodynamic vortex in the physical regime
$N_{\text{p}}>0$. In particular, we shall explicitly prove below
that, in the $N_{\text{p}}>0$ regime of the circulating fluid, the
corresponding acoustic spacetime is characterized by a {\it finite}
discrete set of reflecting cylinder radii,
$\{r_{\text{c}}(N_{\text{p}},m;n)\}^{n=N_{\text{max}}}_{n=1}$, that
can support the marginally-stable static acoustic modes.

\section{Onset of the superradiant instabilities in the polytropic hydrodynamic vortex}

A necessary (though {\it not} a sufficient) condition for the
existence of exponentially growing (superradiantly amplified)
resonant modes in the polytropic hydrodynamic vortex is provided by
the compact inequality \cite{CCP}
\begin{equation}\label{Eq14}
r_{\text{c}}<r_{\text{e}}\  .
\end{equation}
The characteristic inequality (\ref{Eq14}) simply reflects the
physical requirement that the acoustic ergoregion of the effective
$(2+1)$-dimensional rotating spacetime be part of the fluid system.

Intriguingly, however, one finds that not every circulating fluid
system with inner reflecting  boundary conditions at
$r_{\text{c}}<r_{\text{e}}$ is superradiantly unstable to acoustic
perturbation modes \cite{Cars,Hods,CCP}. In particular, for a given
set of the dimensionless physical parameters $\{N_{\text{p}},m\}$
that characterize the polytropic hydrodynamic vortex, there exists a
critical cylinder radius
\begin{equation}\label{Eq15}
x^{*}_{\text{c}}\equiv\text{max}_n\{[x_{\text{c}}(N_{\text{p}},m;n)]^{n=N_{\text{max}}}_{n=1}\}\
\end{equation}
which marks the boundary between stable and superradiantly unstable
circulating fluid configurations. Specifically, circulating fluid
systems with supporting radii in the physical regime
$x_{\text{c}}>x^{*}_{\text{c}}(N_{\text{p}},m)$ are stable to
acoustic perturbation modes with azimuthal harmonic index $m$,
whereas circulating fluid systems with supporting radii in the
physical regime $x_{\text{c}}<x^{*}_{\text{c}}(N_{\text{p}},m)$ are
known to develop superradiant instabilities under sound perturbation
modes with azimuthal harmonic index $m$ \cite{CCP}.

Below we shall explicitly demonstrate that, for certain values of
the dimensionless physical parameters $\{N_{\text{p}},m\}$ which
characterize the polytropic hydrodynamic vortex, the characteristic
{\it discrete} set
$\{x_{\text{c}}(N_{\text{p}},m;n)\}^{n=N_{\text{max}}}_{n=1}$ of
dimensionless cylinder radii that can support the static
(marginally-stable) acoustic resonant modes can be determined {\it
analytically}.

\section{Generic properties of the polytropic hydrodynamic vortex}

In this section we shall discuss three fundamental features of the
polytropic hydrodynamic vortex: (1) the stability properties of the
fundamental $m=1$ sound modes, (2) the finite number of
marginally-stable (static) acoustic modes, and (3) the existence of
a lower bound on the dimensionless polytropic index
$N_{\text{p}}(m)$ of the marginally-stable acoustic resonances.

\subsection{No marginally-stable sound modes for the fundamental
$m=1$ acoustic perturbations}

We shall first prove that there are no marginally-stable (static)
acoustic modes for the fundamental $m=1$ perturbations of the
polytropic hydrodynamic vortex. To this end, we note that the
resonance condition (\ref{Eq13}) takes the remarkably simple form
\begin{equation}\label{Eq16}
{_2F_1}(0,N_{\text{p}}+1;2;x^{-2}_{\text{c}})=0\
\end{equation}
for the $m=1$ acoustic perturbation modes. Using the fact that
${_2F_1}(0,b;c;z)=1$, one immediately deduces from (\ref{Eq16})
that, for the fundamental $m=1$ acoustic perturbations, there are no
static (marginally-stable) resonant modes. Furthermore, remembering
that the static resonances mark the onset of superradiant
instabilities in this physical system \cite{Cars,Hods,CCP}, our
analysis in the present subsection indicates that the polytropic
hydrodynamic vortex is stable to the fundamental $m=1$ perturbation
modes.

\subsection{The number of static (marginally-stable) resonances is
finite}

As emphasized above, it has been proved \cite{Hods} that for
constant density fluids, which are characterized by the simple
relation $N_{\text{p}}=0$ \cite{Cars,Hods}, there is an {\it
infinite} countable set
$\{r_{\text{c}}(N_{\text{p}}=0,m;n)\}^{n=\infty}_{n=1}$ of
reflecting surface radii [with the property
$r_{\text{c}}(n\to\infty)\to0$ \cite{Hods}] that can support the
marginally-stable static acoustic modes.

On the other hand, as we shall now show, the polytropic hydrodynamic
vortex in the physical regime $N_{\text{p}}>0$ is characterized by a
{\it finite} set of reflecting surface radii that can support the
marginally-stable acoustic resonant modes. In particular, we point
out that, for positive integer values of the composed physical
parameter \cite{Notenpm}
\begin{equation}\label{Eq17}
N_{\text{c}}(N_{\text{p}},m)\equiv{{\sqrt{N^2_{\text{p}}+m^2(1+2N_{\text{p}})}-N_{\text{p}}-m}\over{2}}\
,
\end{equation}
the characteristic resonance condition (\ref{Eq13}) is a polynomial
equation of degree $N_{\text{c}}$ in the dimensionless variable
$z\equiv x^{-2}_{\text{c}}$ which yields a {\it finite} number
$N_{\text{c}}$ of dimensionless cylinder radii that can support the
marginally-stable (static) acoustic modes of the polytropic
hydrodynamic vortex.

For non-integer values of the composed physical parameter
$N_{\text{c}}$ [see Eq. (\ref{Eq17})], one can solve numerically the
characteristic resonance condition (\ref{Eq13}) in order to
determine the discrete set of static resonances which characterize
the polytropic hydrodynamic vortex. Doing so, one finds (see Table
\ref{Table1} below) that, for $N_{\text{c}}\not\in\mathbb{N}$ with
$N_{\text{p}}\geq1$, the number $N_{\text{r}}$ of resonances (or
equivalently, the number $N_{\text{r}}$ of supporting cylinder
radii) is given by the simple relation $N_{\text{r}}=\left
\lceil{N_{\text{c}}}\right \rceil$ \cite{Noteceil}.

In addition, taking cognizance of the fact that the supporting radii
of the central reflecting cylinder are characterized by the
dimensionless lower bound [see Eqs. (\ref{Eq10}) and (\ref{Eq12})]
%\cite{Notelc}
\begin{equation}\label{Eq18}
x_{\text{c}}\geq1\  ,
\end{equation}
one may use the characteristic relation \cite{Noteab1}
\begin{equation}\label{Eq19}
{_2F_1}(a,b;c;x_{\text{c}}=1)={{\Gamma(c)\Gamma(c-a-b)}\over{\Gamma(c-a)\Gamma(c-b)}}\
\ \ \ \text{for}\ \ \ \ \Re(c-a-b)>0\
\end{equation}
of the hypergeometric function and the well known pole structure
\cite{Notegc}
\begin{equation}\label{Eq20}
{{1}\over{\Gamma(-n)}}=0\ \ \ \text{for}\ \ \ n=0,1,2,...
\end{equation}
of the Gamma functions to infer from the resonance equation
(\ref{Eq13}) that, in the $N_{\text{p}}<1$ regime \cite{Notenp1} and
for positive integer values of the composed physical quantity
\cite{Notenpb,Notencnc}
\begin{equation}\label{Eq21}
{\bar
N_{\text{c}}}(N_{\text{p}},m)\equiv{{\sqrt{N^2_{\text{p}}+m^2(1+2N_{\text{p}})}+N_{\text{p}}-m}\over{2}}\
,
\end{equation}
a new resonant mode (characterized by the limiting dimensionless
radius $x_{\text{c}}=1$) is added to the discrete family of static
(marginally-stable) resonant modes each time the integer ${\bar
N_{\text{c}}}$ increases by one.

We therefore conclude that, as opposed to the $N_{\text{p}}=0$ case
studied in \cite{Cars,Hods}, the polytropic hydrodynamic vortex in
the physical regime $N_{\text{p}}>0$ is characterized by a {\it
finite} discrete set of dimensionless reflecting radii,
$\{x_{\text{c}}(N_{\text{p}},m;n)\}^{n=N_{\text{r}}}_{n=1}$, that
can support the static (marginally-stable) acoustic resonances,
where \cite{Noteflor,Notewwf,Notex1} [see Eqs. (\ref{Eq17}) and
(\ref{Eq21})]
%\begin{equation}\label{Eq21}
%N_{\text{r}}(N_{\text{p}},m)=
%\begin{cases}
%N_{\text{f}}
%& \text{ if }\ \ \ N_{\text{f}}\in\mathbb{N}\ ; \\
%\lfloor{N_{\text{f}}}\rfloor+1 & \text{ if }\ \ \
%N_{\text{f}}\not\in\mathbb{N}\ \text{and}\ N_{\text{p}}>1\ ; \\
%\lfloor{{\bar N_{\text{f}}}}\rfloor & \text{ if }\ \ \
%N_{\text{f}}\not\in\mathbb{N}\ \text{and}\ N_{\text{p}}<1\  .
%\end{cases}
%\end{equation}
%\begin{equation}\label{Eq21}
%N_{\text{r}}(N_{\text{p}},m)=
%\begin{cases}
%\lfloor{{\bar N_{\text{f}}}}\rfloor & \text{ for }\ \ \ 0<N_{\text{p}}<1\ ; \\
%\lceil{N_{\text{f}}}\rceil & \text{ for }\ \ \ N_{\text{p}}>1\ .
%\end{cases}
%\end{equation}
\begin{equation}\label{Eq22}
N_{\text{r}}(N_{\text{p}},m)=
\begin{cases}
\lfloor{{{\sqrt{N^2_{\text{p}}+m^2(1+2N_{\text{p}})}+N_{\text{p}}-m}\over{2}}}\rfloor &
\text{ for }\ \ \ 0<N_{\text{p}}<1\ ; \\
\lceil{{{\sqrt{N^2_{\text{p}}+m^2(1+2N_{\text{p}})}-N_{\text{p}}-m}\over{2}}}\rceil
& \text{ for }\ \ \ N_{\text{p}}\geq1\ .
\end{cases}
\end{equation}

In Table \ref{Table1} we present the number of reflecting cylinder
radii that can support the static resonant modes of the polytropic
hydrodynamic vortex as deduced by directly solving numerically the
resonance condition (\ref{Eq13}) for various values of the
dimensionless physical parameters $N_{\text{p}}$ and $m$. It is
worth pointing out that the numerical data presented in Table
\ref{Table1} agree with the compact analytical formula (\ref{Eq22})
for the (finite) number of acoustic resonant modes that characterize
the polytropic hydrodynamic vortex in the physical regime
$N_{\text{p}}>0$.

\begin{table}[htbp]
\centering
\begin{tabular}{|c|c|c|}
\hline $\ \ N_{\text{p}}\ \ $\ \ & \ \ $m$ \ \ & \ \ $\#$\ \text{of
resonances}\ \ \\
\hline
\ \ ${1\over2}$\ \ \ &\ \ $1$\ \ &\ \ $0$\ \ \\
\ \ $$\ \ \ &\ \ $2$\ \ &\ \ $0$\ \ \\
\ \ $$\ \ \ &\ \ $3$\ \ &\ \ $0$\ \ \\
\ \ $$\ \ \ &\ \ $4$\ \ &\ \ $1$\ \ \\
\ \ $$\ \ \ &\ \ $5$\ \ &\ \ $1$\ \ \\
%\ \ $3/4$\ \ \ &\ \ $1$\ \ &\ \ $0$\ \ \\
%\ \ $$\ \ \ &\ \ $2$\ \ &\ \ $1$\ \ \\
%\ \ $$\ \ \ &\ \ $3$\ \ &\ \ $1$\ \ \\
%\ \ $$\ \ \ &\ \ $4$\ \ &\ \ $1$\ \ \\
%\ \ $$\ \ \ &\ \ $5$\ \ &\ \ $1$\ \ \\
\ \ $1$\ \ \ &\ \ $1$\ \ &\ \ $0$\ \ \\
\ \ $$\ \ \ &\ \ $2$\ \ &\ \ $1$\ \ \\
\ \ $$\ \ \ &\ \ $3$\ \ &\ \ $1$\ \ \\
\ \ $$\ \ \ &\ \ $4$\ \ &\ \ $1$\ \ \\
\ \ $$\ \ \ &\ \ $5$\ \ &\ \ $2$\ \ \\
\ \ $3$\ \ \ &\ \ $1$\ \ &\ \ $0$\ \ \\
\ \ $$\ \ \ &\ \ $2$\ \ &\ \ $1$\ \ \\
\ \ $$\ \ \ &\ \ $3$\ \ &\ \ $2$\ \ \\
\ \ $$\ \ \ &\ \ $4$\ \ &\ \ $2$\ \ \\
\ \ $$\ \ \ &\ \ $5$\ \ &\ \ $3$\ \ \\
\ \ $5$\ \ \ &\ \ $1$\ \ &\ \ $0$\ \ \\
\ \ $$\ \ \ &\ \ $2$\ \ &\ \ $1$\ \ \\
\ \ $$\ \ \ &\ \ $3$\ \ &\ \ $2$\ \ \\
\ \ $$\ \ \ &\ \ $4$\ \ &\ \ $3$\ \ \\
\ \ $$\ \ \ &\ \ $5$\ \ &\ \ $4$\ \ \\
\hline
\end{tabular}
\caption{Static (marginally-stable) acoustic resonant modes of the
polytropic hydrodynamic vortex. We present, for various values of
the dimensionless physical parameters $N_{\text{p}}$ and $m$, the
number of reflecting cylinder radii that can support the static
resonant modes as computed numerically from the resonance condition
(\ref{Eq13}). The numerically computed data agree with the compact
analytical formula (\ref{Eq22}) which, in the $N_{\text{p}}>0$
regime, determines the (finite) number of static (marginally-stable)
resonant modes that characterize the polytropic hydrodynamic
vortex.} \label{Table1}
\end{table}

It is interesting to stress the fact that our results in the present
subsection reveal the fact that the $N_{\text{p}}\to0$ limit of the
polytropic hydrodynamic vortex is not continuous. In particular,
while the $N_{\text{p}}=0$ case of constant density fluids studied
in \cite{Cars,Hods} is characterized, for every positive value of
the azimuthal harmonic index $m\geq1$, by an {\it infinite}
countable set of marginally-stable acoustic resonances \cite{Hods},
in the $N_{\text{p}}\to 0^+$ limit of the polytropic hydrodynamic
vortex and for a fixed value of the integer field parameter $m$
there are {\it no} static (marginally-stable) acoustic resonant
modes \cite{Notenum}.

\subsection{Lower bound on the polytropic index $N_{\text{p}}(m)$ of
the marginally-stable acoustic resonances}

From the analytical formula (\ref{Eq22}) for the number
$N_{\text{r}}(N_{\text{p}},m)$ of marginally-stable acoustic
resonant modes one deduces that, for a given value of the azimuthal
harmonic index $m$, the static resonant modes of the polytropic
hydrodynamic vortex can only exist in the regime
\cite{Notelst,Notelt1}
\begin{equation}\label{Eq23}
N_{\text{p}}(m)\geq
N^{\text{min}}_{\text{p}}(m)\equiv{{2(m+1)}\over{m(m+1)+2}}\ .
\end{equation}

Equivalently, the inequality (\ref{Eq23}) provides, for a given
value of the dimensionless polytropic index $N_{\text{p}}$ of the
circulating fluid, the lower bound \cite{Notelcc,Noteminn}
\begin{equation}\label{Eq24}
m(N_{\text{p}})\geq m^{\text{min}}(N_{\text{p}})\equiv
\begin{cases}
\lceil{{{2-N_{\text{p}}+\sqrt{4+4N_{\text{p}}-7N^2_{\text{p}}}}\over{2N_{\text{p}}}}}\rceil
& \text{ for } \ \ \ 0<N_{\text{p}}<{3\over4}\ ; \\
2 & \text{ for }\ \ \ N_{\text{p}}\geq{3\over4}\
\end{cases}
\end{equation}
on the values of the azimuthal harmonic indices that can trigger
rotational instabilities in the polytropic hydrodynamic vortex.
Interestingly, from the relation (\ref{Eq24}) one finds the simple
asymptotic behavior
\begin{equation}\label{Eq25}
m^{\text{min}}(N_{\text{p}}\to0)\to\infty\  ,
\end{equation}
which implies that only acoustic modes with large values
[$m(N_{\text{p}}\ll1)\geq \lceil{2/N_{\text{p}}}\rceil\gg1]$ of the
azimuthal harmonic index $m$ can trigger rotational instabilities in
the $N_{\text{p}}\ll1$ regime.

\section{Upper bound on the supporting radii of the marginally-stable (static) resonant modes}

In the present section we shall explicitly prove that, for given
dimensionless physical parameters $\{N_{\text{p}},m\}$, one can
derive an upper bound [which is {\it stronger} than the bound
(\ref{Eq14})] on the inner supporting radii
$\{x_{\text{c}}(N_{\text{p}},m;n)\}$ which characterize the
marginally-stable resonant modes of the polytropic hydrodynamic
vortex.

To this end, we shall first write the radial differential equation
(\ref{Eq7}) in the form
\begin{equation}\label{Eq26}
x^2(x^2-1){{d^2\Phi}\over{dx^2}}+\big[2\alpha
x(x^2-1)+2N_{\text{p}}x\big]{{d\Phi}\over{dx}}+
\big[(\alpha^2-\alpha-m^2+{1\over4})(x^2-1)+N_{\text{p}}(2\alpha-1+2m^2)\big]\Phi=0\
,
\end{equation}
where
\begin{equation}\label{Eq27}
\psi(x)=x^{\alpha}\cdot\Phi(x)\ \ \ \text{with}\ \ \ \alpha>1/2-m\
.
\end{equation}
Taking cognizance of the inner boundary condition (\ref{Eq11}) at
the surface of the reflecting cylinder and the characteristic
asymptotic behavior $\psi(x\to\infty)\to A\cdot x^{1/2-m}$ of the
radial eigenfunction (\ref{Eq9}) \cite{Notepa}, one deduces that the
radial function $\Phi(x)$, which characterizes the marginally-stable
(static) resonances of the polytropic hydrodynamic vortex, must have
(at least) one extremum point, $x=x_{\text{peak}}$, with the
properties \cite{Noteinf}
\begin{equation}\label{Eq28}
\{\Phi\neq0\ \ \ ; \ \ \ {{d\Phi}\over{dx}}=0\ \ \ ; \ \ \
\Phi{{d^2\Phi}\over{dx^2}}<0\}\ \ \ \ \text{for}\ \ \ \
x=x_{\text{peak}}\ ,
\end{equation}
where
\begin{equation}\label{Eq29}
x_{\text{peak}}\in (x_{\text{c}},\infty)\  .
\end{equation}
Substituting the functional relations (\ref{Eq28}) of the radial
eigenfunction $\Phi(x)$ into Eq. (\ref{Eq26}), one obtains the
characteristic inequality
\begin{equation}\label{Eq30}
(\alpha^2-\alpha-m^2+{1\over4})(x^2_{\text{peak}}-1)+N_{\text{p}}(2\alpha-1+2m^2)>0\
\end{equation}
at the extremum point (\ref{Eq29}).

Assuming
\begin{equation}\label{Eq31}
\alpha(\alpha-1)<m^2-{1\over4}\  ,
\end{equation}
one finds from (\ref{Eq30}) the inequality
\begin{equation}\label{Eq32}
x^2_{\text{peak}}<1+N_{\text{p}}\cdot
{{2m^2-1+2\alpha}\over{m^2-{1\over4}-\alpha(\alpha-1)}}\  .
\end{equation}
The strongest upper bound on the radial location of the extremum
point $x=x_{\text{peak}}$ can be obtained by minimizing, with
respect to the dimensionless parameter $\alpha$, the functional
expression on the right-hand-side of (\ref{Eq32}). In particular,
this expression is minimized for \cite{Noteres}
\begin{equation}\label{Eq33}
\alpha(m)=\alpha^*(m)={1\over 2}-m^2+m\sqrt{m^2-1}\  ,
\end{equation}
in which case one finds from Eqs. (\ref{Eq29}), (\ref{Eq32}), and
(\ref{Eq33}) the upper bound
\begin{equation}\label{Eq34}
x_{\text{c}}<\sqrt{1+N_{\text{p}}\cdot\Big(1+\sqrt{1-{{1}\over{m^2}}}\Big)}\
\end{equation}
on the characteristic inner supporting radii of the polytropic
hydrodynamic vortex.

It is worth noting that, taking cognizance of the characteristic
relation (\ref{Eq3}) and defining the dimensionless radial
coordinate
\begin{equation}\label{Eq35}
{\bar x_{\text{c}}}\equiv {{r_{\text{c}}}\over{r_{\text{e}}}}\  ,
\end{equation}
one can express the analytically derived upper bound (\ref{Eq34}) in
the form
\begin{equation}\label{Eq36}
{\bar
x_{\text{c}}}<\sqrt{{{1+N_{\text{p}}\cdot\Big(1+\sqrt{1-{{1}\over{m^2}}}\Big)}\over{1+2N_{\text{p}}}}}\
.
\end{equation}

\section{Analytic treatment of the resonance condition
for marginally-stable resonant modes in the small-radii regime}

In the present section we shall explicitly show that the resonance
equation (\ref{Eq13}), which determines the unique family
$\{x_{\text{c}}(N_{\text{p}},m;n)\}$ of dimensionless supporting
radii that characterize the static (marginally-stable) resonant
modes of the polytropic hydrodynamic vortex, is amenable to an {\it
analytical} treatment in the small-radii regime
\begin{equation}\label{Eq37}
%x_{\text{c}}\to1^+\  .
x_{\text{c}}-1\ll1\  .
\end{equation}

Taking cognizance of the characteristic limiting behavior [see Eq. 15.1.1 of \cite{Abram}]
\begin{equation}\label{Eq38}
{_2F_1}(a,b;c;z\to0)\to1\
\end{equation}
of the hypergeometric function and using Eqs. 6.1.15 and 15.3.6 of
\cite{Abram}, one can express the resonance condition (\ref{Eq13})
in the form
%\begin{equation}\label{Eq36}
%(1-x^{-2}_{\text{c}})^{1-N_{\text{p}}}=-{{\Gamma(1-N_{\text{p}})
%\Gamma\Big[{{N_{\text{p}}-\sqrt{N^2_{\text{p}}+m^2(1+2N_{\text{p}})}+m}\over{2}}\Big]
%\Gamma\Big[{{N_{\text{p}}+\sqrt{N^2_{\text{p}}+m^2(1+2N_{\text{p}})}+m}\over{2}}\Big]}
%\over
%{\Gamma(N_{\text{p}}-1)\Gamma\Big[1-{{N_{\text{p}}-\sqrt{N^2_{\text{p}}+m^2(1+2N_{\text{p}})}-m}\over{2}}\Big]
%\Gamma\Big[1-{{N_{\text{p}}+\sqrt{N^2_{\text{p}}+m^2(1+2N_{\text{p}})}-m}\over{2}}\Big]}}
%\ \ \ \ \text{for}\ \ \ \ x_{\text{c}}\to1^+\  .
%\end{equation}
%Using Eq. 6.1.15 of \cite{Abram}, one can write the resonance
%\begin{equation}\label{Eq36}
%(1-x^{-2}_{\text{c}})^{1-N_{\text{p}}}={{2\Gamma(1-N_{\text{p}})
%\Gamma\Big[{{N_{\text{p}}-\sqrt{N^2_{\text{p}}+m^2(1+2N_{\text{p}})}+m}\over{2}}\Big]
%\Gamma\Big[{{N_{\text{p}}+\sqrt{N^2_{\text{p}}+m^2(1+2N_{\text{p}})}+m}\over{2}}\Big]}
%\over
%{m(m+1)N_{\text{p}}\Gamma(N_{\text{p}}-1)\Gamma\Big[-{{N_{\text{p}}-\sqrt{N^2_{\text{p}}+m^2(1+2N_{\text{p}})}-m}\over{2}}\Big]
%\Gamma\Big[-{{N_{\text{p}}+\sqrt{N^2_{\text{p}}+m^2(1+2N_{\text{p}})}-m}\over{2}}\Big]}}
%\ \ \ \ \text{for}\ \ \ \ x_{\text{c}}\to1^+\  .
%\end{equation}
\begin{equation}\label{Eq39}
(1-x^{-2}_{\text{c}})^{1-N_{\text{p}}}={{2(1-N_{\text{p}})\Gamma(-N_{\text{p}})
\Gamma\Big[{{N_{\text{p}}-\sqrt{N^2_{\text{p}}+m^2(1+2N_{\text{p}})}+m}\over{2}}\Big]
\Gamma\Big[{{N_{\text{p}}+\sqrt{N^2_{\text{p}}+m^2(1+2N_{\text{p}})}+m}\over{2}}\Big]}
\over
{m(m+1)\Gamma(N_{\text{p}})\Gamma\Big[-{{N_{\text{p}}-\sqrt{N^2_{\text{p}}+m^2(1+2N_{\text{p}})}-m}\over{2}}\Big]
\Gamma\Big[-{{N_{\text{p}}+\sqrt{N^2_{\text{p}}+m^2(1+2N_{\text{p}})}-m}\over{2}}\Big]}}
\ \ \ \ \text{for}\ \ \ \ x_{\text{c}}\to1^+\  .
\end{equation}

\subsection{The $0<N_{\text{p}}<1$ regime}

We shall first solve the analytically derived resonance equation
(\ref{Eq39}) in the $0<N_{\text{p}}<1$ regime, which corresponds to
the strong inequality $(1-x^{-2}_{\text{c}})^{1-N_{\text{p}}}\ll1$
[see Eqs. (\ref{Eq37}) and (\ref{Eq39})]. Substituting into
(\ref{Eq39})
\begin{equation}\label{Eq40}
N_{\text{p}}(m;n)=N^{*}_{\text{p}}\times(1+\epsilon)\ \ \ \
\text{with}\ \ \ \ \epsilon\ll1\  ,
\end{equation}
where
\begin{equation}\label{Eq41}
N^{*}_{\text{p}}(m;n)={{2n(m+n)}\over{m(m+1)+2n}}\ \ \ \ ; \ \ \ \
n=1,2,3,...\ ,
\end{equation}
one obtains the linearized (small-$\epsilon$) resonance condition
\cite{Notesmx}
\begin{equation}\label{Eq42}
(1-x^{-2}_{\text{c}})^{1-N^{*}_{\text{p}}}={{2\alpha(-1)^n
n!(1-N^{*}_{\text{p}})\Gamma(-N^{*}_{\text{p}})
\Gamma\big[N^{*}_{\text{p}}-\sqrt{N^{*2}_{\text{p}}+m^2(1+2N^{*}_{\text{p}})}-n\big]
\Gamma\big(N^{*}_{\text{p}}-n\big)} \over
{m(m+1)\Gamma(N^{*}_{\text{p}})\Gamma\big[-\sqrt{N^{*2}_{\text{p}}+m^2(1+2N^{*}_{\text{p}})}-n\big]}}
\times\epsilon+O(\epsilon^2)\  ,
% \ \ \ \ \text{for}\ \ \ \ x_{\text{c}}\to1^+\  ,
\end{equation}
%\begin{equation}\label{Eq36}
%(1-x^{-2}_{\text{c}})^{1-N_{\text{p}}}={{2(1-N_{\text{p}})\Gamma(-N_{\text{p}})
%\Gamma\big[-{{m(m-1)n}\over{m(m+1)+2n}}\big]
%\Gamma\big(m+n\big)}
%\over
%{m(m+1)\Gamma(N_{\text{p}})\Gamma\big[{{m(m+1)(m+n)}\over{m(m+1)+2n}}\big]
%\Gamma\big[-n(1+\epsilon)\big]}}
%\ \ \ \ \text{for}\ \ \ \ x_{\text{c}}\to1^+\  .
%\end{equation}
where
\begin{equation}\label{Eq43}
\alpha\equiv {1\over2}\Big[{{N^{*}_{\text{p}}(N^{*}_{\text{p}}+m^2)}
\over{\sqrt{N^{*2}_{\text{p}}+m^2(1+2N^{*}_{\text{p}})}}}-N^{*}_{\text{p}}\Big]\
.
\end{equation}
From (\ref{Eq42}) one finds the resonant solution
\begin{equation}\label{Eq44}
x_{\text{c}}(\epsilon,m;n)=1+\beta_1\times\epsilon^{{1}\over{{1-N^{*}_{\text{p}}}}}\
\ \ \ \text{for}\ \ \ \ 0<N_{\text{p}}<1\
\end{equation}
in the $\epsilon\ll1$ regime [or equivalently, in the
$x_{\text{c}}-1\ll1$ regime, see (\ref{Eq37})], where
\begin{equation}\label{Eq45}
\beta_1\equiv
{1\over2}\Bigg\{{{\big[{{N^{*}_{\text{p}}(N^{*}_{\text{p}}+m^2)}
\over{\sqrt{N^{*2}_{\text{p}}+m^2(1+2N^{*}_{\text{p}})}}}-N^{*}_{\text{p}}\big](-1)^n
n!(1-N^{*}_{\text{p}})\Gamma(-N^{*}_{\text{p}})
\Gamma\big[N^{*}_{\text{p}}-\sqrt{N^{*2}_{\text{p}}+m^2(1+2N^{*}_{\text{p}})}-n\big]
\Gamma\big(N^{*}_{\text{p}}-n\big)} \over
{m(m+1)\Gamma(N^{*}_{\text{p}})\Gamma\big[-\sqrt{N^{*2}_{\text{p}}+m^2(1+2N^{*}_{\text{p}})}-n\big]}}\Bigg\}
^{{1}\over{{1-N^{*}_{\text{p}}}}}\  .
\end{equation}

\subsection{The $N_{\text{p}}>1$ regime}

We shall next solve the resonance equation (\ref{Eq39}) in the
$N_{\text{p}}>1$ regime, which corresponds to the strong inequality
$(1-x^{-2}_{\text{c}})^{1-N_{\text{p}}}\gg1$ [see Eqs. (\ref{Eq37})
and (\ref{Eq39})]. Substituting into (\ref{Eq39})
\begin{equation}\label{Eq46}
N_{\text{p}}(m;n)=N^{\#}_{\text{p}}\times(1+\epsilon)\ \ \ \
\text{with}\ \ \ \ 0<\epsilon\ll1\  ,
\end{equation}
where \cite{Notens}
\begin{equation}\label{Eq47}
N^{\#}_{\text{p}}(m;n)={{2n(m+n)}\over{m(m-1)-2n}}\ \ \ \ ; \ \ \ \
n=1,2,3,...\ ,
\end{equation}
one obtains the linearized (small-$\epsilon$) resonance condition
\cite{Notesmx2}
\begin{equation}\label{Eq48}
(1-x^{-2}_{\text{c}})^{1-N^{\#}_{\text{p}}}={{2(1-N^{\#}_{\text{p}})\Gamma(-N^{\#}_{\text{p}})
\Gamma\big[\sqrt{N^{\#2}_{\text{p}}+m^2(1+2N^{*}_{\text{p}})}-n\big]}
\over {{\bar\alpha}m(m+1)(-1)^n n!\Gamma(N^{\#}_{\text{p}})
\Gamma\big[-N^{\#}_{\text{p}}+\sqrt{N^{\#2}_{\text{p}}+m^2(1+2N^{\#}_{\text{p}})}-n\big]
\Gamma(-N^{\#}_{\text{p}}-n)}} \times\epsilon^{-1}+O(1)\  ,
\end{equation}
where
\begin{equation}\label{Eq49}
{\bar\alpha}\equiv
-{1\over2}\Big[{{N^{\#}_{\text{p}}(N^{\#}_{\text{p}}+m^2)}
\over{\sqrt{N^{\#2}_{\text{p}}+m^2(1+2N^{\#}_{\text{p}})}}}-N^{\#}_{\text{p}}\Big]\
.
\end{equation}
From (\ref{Eq48}) one finds the resonant solution
\begin{equation}\label{Eq50}
x_{\text{c}}(\epsilon,m;n)=1+\beta_2\times\epsilon^{{1}\over{{N^{\#}_{\text{p}}-1}}}\
\ \ \ \text{for}\ \ \ \ N_{\text{p}}>1\
\end{equation}
in the $\epsilon\ll1$ regime, where
\begin{equation}\label{Eq51}
\beta_2\equiv
{1\over2}\Bigg\{{{4(N^{\#}_{\text{p}}-1)\Gamma(-N^{\#}_{\text{p}})
\Gamma\big[\sqrt{N^{\#2}_{\text{p}}+m^2(1+2N^{*}_{\text{p}})}-n\big]}
\over {m(m+1)(-1)^n
n!\big[{{N^{\#}_{\text{p}}(N^{\#}_{\text{p}}+m^2)}
\over{\sqrt{N^{\#2}_{\text{p}}+m^2(1+2N^{\#}_{\text{p}})}}}-N^{\#}_{\text{p}}\big]\Gamma(N^{\#}_{\text{p}})
\Gamma\big[-N^{\#}_{\text{p}}+\sqrt{N^{\#2}_{\text{p}}+m^2(1+2N^{\#}_{\text{p}})}-n\big]
\Gamma(-N^{\#}_{\text{p}}-n)}}\Bigg\}
^{{1}\over{{1-N^{\#}_{\text{p}}}}}\  .
\end{equation}

%\subsection{Cases in which $N_{\text{p}}$ is nearly an integer}
%
%As we shall now show, the regime $x_{\text{c}}-1\ll1$ [see
%(\ref{Eq37})] is also covered by cases in which the dimensionless
%polytropic index $N_{\text{p}}$ of the circulating fluid is nearly
%an {\it integer}. In particular, substituting into (\ref{Eq39})
%\begin{equation}\label{Eq52}
%N_{\text{p}}=n\times(1+\epsilon)\ \ \ \ \text{with}\ \ \ \
%\{\epsilon\ll1\ \ \ \text{and}\ \ \ n=2,3,4,...\}\  ,
%\end{equation}
%one obtains the linearized (small-$\epsilon$) resonance equation
%\cite{Notesmx3}
%\begin{equation}\label{Eq53}
%(1-x^{-2}_{\text{c}})^{1-n}={{2(n-1)
%\Gamma\Big[{{n-\sqrt{n^2+m^2(1+2n)}+m}\over{2}}\Big]
%\Gamma\Big[{{n+\sqrt{n^2+m^2(1+2n)}+m}\over{2}}\Big]} \over
%{m(m+1)(-1)^{n}(n!)^2\Gamma\Big[-{{n-\sqrt{n^2+m^2(1+2n)}-m}\over{2}}\Big]
%\Gamma\Big[-{{n+\sqrt{n^2+m^2(1+2n)}-m}\over{2}}\Big]}}\times
%\epsilon^{-1}+O(1)\ .
%\end{equation}
%From (\ref{Eq53}) one finds the resonant solution
%\begin{equation}\label{Eq54}
%x_{\text{c}}(\epsilon,m;n)=1+\beta_3\times\epsilon^{{1}\over{{n-1}}}\
%\ \ \ \text{for}\ \ \ \ N_{\text{p}}\simeq n=2,3,4,...\
%\end{equation}
%in the $\epsilon\ll1$ regime [or equivalently, in the
%$x_{\text{c}}-1\ll1$ regime, see (\ref{Eq37})], where
%\begin{equation}\label{Eq55}
%\beta_3\equiv {1\over2}\Bigg\{{{2(n-1)
%\Gamma\Big[{{n-\sqrt{n^2+m^2(1+2n)}+m}\over{2}}\Big]
%\Gamma\Big[{{n+\sqrt{n^2+m^2(1+2n)}+m}\over{2}}\Big]} \over
%{m(m+1)(-1)^{n}(n!)^2\Gamma\Big[-{{n-\sqrt{n^2+m^2(1+2n)}-m}\over{2}}\Big]
%\Gamma\Big[-{{n+\sqrt{n^2+m^2(1+2n)}-m}\over{2}}\Big]}}\Bigg\}
%^{{1}\over{{1-n}}}\ .
%\end{equation}

It is interesting to stress the fact that, taking cognizance of the
analytically derived resonant formulas (\ref{Eq44}) and
(\ref{Eq50}), one learns that, in general, the characteristic
supporting radii $\{x_{\text{c}}(\epsilon,m;n)\}$ of the polytropic
hydrodynamic vortex have a non-trivial ({\it non}-linear) dependence
on the dimensionless small parameter $\epsilon$.

\section{Analytic treatment of the resonance condition:
monotonic behavior of the outermost supporting radii ${\bar
x^{\text{max}}_{\text{c}}}$}

The resonance equation (\ref{Eq13}), which determines the static
(marginally-stable) resonances of the polytropic hydrodynamic
vortex, can be solved numerically. In particular, using numerical
techniques, it was nicely demonstrated in \cite{CCP} that the
largest (outermost) dimensionless supporting radius ${\bar
x^{\text{max}}_{\text{c}}}(N_{\text{p}},m)$ [see Eq. (\ref{Eq35})]
of the central reflecting cylinder increases monotonically with
increasing values of the azimuthal harmonic index $m$ and decreasing
values of the dimensionless polytropic index $N_{\text{p}}$ of the
circulating fluid. It is worth emphasizing again that the physical
significance of the quantity ${\bar
x^{\text{max}}_{\text{c}}}(N_{\text{p}},m)$ stems from the fact
that, for given dimensionless physical parameters
$\{N_{\text{p}},m\}$, this outermost supporting radius marks the
boundary between stable and superradiantly unstable configurations
of the polytropic hydrodynamic vortex.

In the present section we shall provide an {\it analytical}
explanation for the interesting monotonic behavior, first observed
numerically in \cite{CCP}, which characterizes the outermost
supporting radius ${\bar x^{\text{max}}_{\text{c}}}(N_{\text{p}},m)$
of the marginally-stable (static) resonant modes of the polytropic
hydrodynamic vortex. In particular, we shall use analytical
techniques in order to demonstrate that the characteristic
supporting radius ${\bar x^{\text{max}}_{\text{c}}}$ increases
monotonically with increasing values of the azimuthal harmonic index
$m$ and decreasing values of the dimensionless polytropic index
$N_{\text{p}}$.

Interestingly, as we shall now demonstrate explicitly, the resonance
condition (\ref{Eq13}), which determines the discrete family
$\{x_{\text{c}}(N_{\text{p}},m;n)\}$ of inner supporting radii that
characterize the marginally-stable resonant modes of the polytropic
hydrodynamic vortex, is amenable to an {\it analytical} treatment in
cases where
\begin{equation}\label{Eq52}
{1\over2}\big[N_{\text{p}}-\sqrt{N^2_{\text{p}}+m^2(1+2N_{\text{p}})}+m\big]=-n\
\ \ \ ; \ \ \ \ n=0,1,2,...\ .
\end{equation}
In particular, in these cases, which correspond to the relation
\begin{equation}\label{Eq53}
N_{\text{p}}(m;n)={{2n(m+n)}\over{m(m-1)-2n}}\  ,
\end{equation}
one finds that the resonance condition (\ref{Eq13}) becomes a
polynomial equation of degree $n$ in the dimensionless physical
variable
\begin{equation}\label{Eq54}
z_{\text{c}}\equiv {{1}\over{x^2_{\text{c}}}}\  .
\end{equation}

For example, in the first nontrivial \cite{Noten0} case, $n=1$, the
resonance condition (\ref{Eq13}) yields the simple linear equation
\begin{equation}\label{Eq55}
{{m(1-m)\cdot z_{\text{c}}+m(m-1)-2}\over{m(m-1)-2}}=0\ \ \ \ ; \ \
\ \ m\geq3\
\end{equation}
for the variable $z_{\text{c}}$. Substituting (\ref{Eq54}) into
(\ref{Eq55}), one finds the simple analytical expression
\cite{Noteok1}
\begin{equation}\label{Eq56}
x_{\text{c}}=\sqrt{{{m(m-1)}\over{(m-2)(m+1)}}}\ \ \ \ ; \ \ \ \
N_{\text{p}}={{2}\over{m-2}}\ \ \ \text{with}\ \ \ m\geq3\ .
\end{equation}
Taking cognizance of the characteristic relation (\ref{Eq3}), one
finds from (\ref{Eq56}) the compact expression
\begin{equation}\label{Eq57}
{\bar x_{\text{c}}}=\sqrt{{{m(m-1)}\over{(m+1)(m+2)}}}\ \ \ \ ; \ \
\ \ N_{\text{p}}={{2}\over{m-2}}\ \ \ \text{with}\ \ \ m\geq3\ .
\end{equation}

As another analytically solvable example, let us consider the $n=2$
case [see Eq. (\ref{Eq53})], in which case the resonance condition
(\ref{Eq13}) yields the quadratic equation
\begin{equation}\label{Eq58}
{{(m^5+m^4-5m^3-m^2+4m)\cdot
z^2_{\text{c}}+(-2m^5+14m^3+4m^2-16m)\cdot
z_{\text{c}}+(m^5-m^4-9m^3+m^2+24m+16)}\over{(m+1)(m^2-m-4)^2}}=0\
\end{equation}
for the dimensionless variable $z_{\text{c}}$. Substituting
(\ref{Eq54}) into (\ref{Eq58}), one obtains the two supporting radii
\cite{Notem3,Noteok2}
\begin{equation}\label{Eq59}
x^{\pm}_{\text{c}}=\sqrt{{{m(m-1)(m+1)(m^2+m-4)}\over{(m^2-m-4)[m(m-1)(m+2)\pm\sqrt{m(m-1)(m^2+3m+4)}]}}}\
\ \ \ ; \ \ \ \ N_{\text{p}}={{4(m+2)}\over{m(m-1)-4}}\ \ \
\text{with}\ \ \ m\geq3\  .
\end{equation}
Taking cognizance of the dimensionless ratio (\ref{Eq3}), one finds
from (\ref{Eq59}) the analytical expression
\begin{equation}\label{Eq60}
{\bar
x^{\pm}_{\text{c}}}=\sqrt{{{m(m-1)(m+1)(m^2+m-4)}\over{(m+3)(m+4)[m(m-1)(m+2)\pm\sqrt{m(m-1)(m^2+3m+4)}]}}}\
\ \ \ ; \ \ \ \ N_{\text{p}}={{4(m+2)}\over{m(m-1)-4}}\ \ \
\text{with}\ \ \ m\geq3\
\end{equation}
for the dimensionless supporting radii of the polytropic
hydrodynamic vortex.

It is physically interesting to point out that the analytically
derived expressions (\ref{Eq57}) and (\ref{Eq60}) reveal the fact
that the dimensionless supporting radius ${\bar
x^{\text{max}}_{\text{c}}}$, which characterizes the
marginally-stable static ($\omega=0$) resonances of the polytropic
hydrodynamic vortex, increases monotonically with increasing values
of the azimuthal harmonic index $m$ and decreasing values
\cite{Noteppm} of the dimensionless polytropic index $N_{\text{p}}$
of the circulating fluid. It is important to stress the fact that
this {\it analytically} demonstrated monotonic behavior of the
characteristic supporting radius ${\bar x^{\text{max}}_{\text{c}}}$
agrees with the {\it numerical} results recently presented in the
interesting work of Oliveira, Cardoso, and Crispino \cite{CCP}.

\section{Summary and discussion}

Contrary to the asymptotically flat spinning Kerr black-hole
spacetime, whose stability to massless bosonic perturbations relies
on the characteristic absorptive properties of its horizon
\cite{PressTeu2,Whit,Cars}, horizonless rotating spacetimes with
ergoregions and inner reflecting boundary conditions may develop
superradiant instabilities to co-rotating integer-spin (bosonic)
fields \cite{Fri,Cars,Carsup,Sla}. In particular, it has recently
been demonstrated that the polytropic hydrodynamic vortex
\cite{CCP}, an effective $(2+1)$-dimensional acoustic spacetime with
an ergoregion of radius $r_{\text{e}}$ and an inner reflecting
boundary at $r=r_{\text{c}}$ may develop exponentially growing
superradiant instabilities \cite{Cars,Hods,CCP}.

In the present paper we have used analytical techniques in order to
explore the physical and mathematical properties of the static
acoustic resonances which characterize the polytropic hydrodynamic
vortex. The physical significance of these marginally-stable sound
modes stems from the fact that these static resonances mark the
boundary between stable and superradiantly unstable configurations
of the effective $(2+1)$-dimensional circulating fluid system
\cite{Cars,Hods,CCP}. The main results derived in this paper and
their physical implications are:

(1) It has been explicitly proved that, for a given value of the
azimuthal harmonic index $m$, the marginally-stable (static)
resonant modes of the polytropic hydrodynamic vortex are restricted
to the dimensionless physical regime [see Eq. (\ref{Eq23})]
\begin{equation}\label{Eq61}
N_{\text{p}}(m)\geq N^{\text{min}}_{\text{p}}\ \ \ \ \text{with}\ \
\ \ N^{\text{min}}_{\text{p}}={{2(m+1)}\over{m(m+1)+2}}\ .
\end{equation}
This characteristic inequality implies, in particular, that in the
$0<N_{\text{p}}\ll1$ regime, only acoustic resonant modes with large
azimuthal indices [see Eq. (\ref{Eq24})],
\begin{equation}\label{Eq62}
m\geq \lceil{2/N_{\text{p}}}\rceil\gg1\  ,
\end{equation}
can trigger superradiant instabilities in the polytropic
hydrodynamic vortex.

(2) We have proved that, for a given set $\{N_{\text{p}},m\}$ of the
dimensionless physical parameters that characterize the polytropic
hydrodynamic vortex, the compact inequality [see Eqs. (\ref{Eq12})
and (\ref{Eq34})]
\begin{equation}\label{Eq63}
{{r_{\text{c}}}\over{r_{\text{s}}}}<\sqrt{1+N_{\text{p}}\cdot\Big(1+\sqrt{1-{{1}\over{m^2}}}\Big)}\
\end{equation}
provides an upper bound on the characteristic supporting radii of
the marginally-stable (static) resonant modes of the system.

(3) It has been shown that the polytropic hydrodynamic vortex is
characterized by a {\it finite} unique family of dimensionless
cylinder radii,
$\{x_{\text{c}}(N_{\text{p}},m;n)\}^{n=N_{\text{r}}}_{n=1}$, that
can support the marginally-stable (static) acoustic resonant modes.
In particular, in the $N_{\text{p}}>0$ regime, one finds
\cite{Noteceil,Noteflor,Notex1} [see Eq. (\ref{Eq22})]
\begin{equation}\label{Eq64}
N_{\text{r}}(N_{\text{p}},m)=
\begin{cases}
\lfloor{{{\sqrt{N^2_{\text{p}}+m^2(1+2N_{\text{p}})}+N_{\text{p}}-m}\over{2}}}\rfloor
&
\text{ for }\ \ \ 0<N_{\text{p}}<1\ ; \\
\lceil{{{\sqrt{N^2_{\text{p}}+m^2(1+2N_{\text{p}})}-N_{\text{p}}-m}\over{2}}}\rceil
& \text{ for }\ \ \ N_{\text{p}}\geq1\ .
\end{cases}
\end{equation}

(4) Interestingly, the fact that the polytropic hydrodynamic vortex
in the physical regime $N_{\text{p}}>0$ is characterized by a {\it
finite} discrete set
$\{x_{\text{c}}(N_{\text{p}},m;n)\}^{n=N_{\text{r}}}_{n=1}$ of
dimensionless cylinder radii that can support the marginally-stable
(static) resonant modes [see Eq. (\ref{Eq64})] should be contrasted
with the $N_{\text{p}}=0$ case studied in \cite{Cars,Hods}. In
particular, it has been explicitly proved in \cite{Hods} that, for
the $N_{\text{p}}=0$ case, the effective $(2+1)$-dimensional
acoustic spacetime is characterized by an {\it infinite} countable
set of cylinder radii,
$\{r_{\text{c}}(N_{\text{p}}=0,m;n)\}^{n=\infty}_{n=1}$, that can
support the marginally-stable resonant sound modes.

(5) Our analysis has revealed the intriguing fact that the
$N_{\text{p}}\to0$ limit of the polytropic hydrodynamic vortex is
not continuous. In particular, while the $N_{\text{p}}=0$ case of
constant density fluids studied in \cite{Cars,Hods} is characterized
by an {\it infinite} countable set of marginally-stable (static)
resonant modes \cite{Hods}, in the $N_{\text{p}}\to 0^+$ limit of
the polytropic hydrodynamic vortex and for a fixed value of the
azimuthal harmonic index $m$ there are {\it no} marginally-stable
acoustic resonant modes \cite{Notenum}.

(6) It has been explicitly shown that the resonance condition
(\ref{Eq13}), which determines the static (marginally-stable)
resonant modes of the polytropic hydrodynamic vortex, can be solved
{\it analytically} in the dimensionless regime
\begin{equation}\label{Eq65}
x_{\text{c}}-1\ll1\
\end{equation}
of small supporting radii. In particular, we have revealed the
intriguing fact that, in the small-radii regime (\ref{Eq65}), the
supporting radii $\{x_{\text{c}}(\epsilon,m;n)\}$ of the polytropic
hydrodynamic vortex are characterized by the non-trivial ({\it
non}-linear) functional dependence [see Eqs. (\ref{Eq44}) and
(\ref{Eq50})]
\begin{equation}\label{Eq66}
x_{\text{c}}(\epsilon)=1+\beta\times\epsilon^{{1}\over{|1-N_{\text{p}}|}}\
,
\end{equation}
where the dimensionless deviation parameter $\epsilon$ is defined in
(\ref{Eq40}) and (\ref{Eq46}).

(7) We have explicitly shown that the characteristic resonance
equation (\ref{Eq13}) of the polytropic hydrodynamic vortex can be
solved {\it analytically} in the dimensionless regime
\begin{equation}\label{Eq67}
N_{\text{p}}(m;n)={{2n(m+n)}\over{m(m-1)-2n}}\ \ \ ; \ \ \
n=0,1,2,...\  .
\end{equation}
In particular, using analytical techniques, we have explicitly
demonstrated that the dimensionless supporting radius ${\bar
x^{\text{max}}_{\text{c}}}$ [see Eq. (\ref{Eq35})], which
characterizes the marginally-stable resonant modes of the polytropic
hydrodynamic vortex, increases monotonically with increasing values
of the azimuthal harmonic index $m$ and decreasing values
\cite{Noteppm} of the dimensionless polytropic index $N_{\text{p}}$
of the circulating fluid [see Eqs. (\ref{Eq57}) and (\ref{Eq60})].
It is important to stress the fact that the characteristic monotonic
behavior of the outermost supporting radius ${\bar
x^{\text{max}}_{\text{c}}(N_{\text{p}},m)}$, which has been
explicitly demonstrated {\it analytically} in the present paper,
agrees with the interesting {\it numerical} results presented
recently in \cite{CCP} for the polytropic hydrodynamic vortex.

\bigskip
\noindent
{\bf ACKNOWLEDGMENTS}
\bigskip

This research is supported by the Carmel Science Foundation. I thank
Yael Oren, Arbel M. Ongo, Ayelet B. Lata, and Alona B. Tea for
stimulating discussions.

%\newpage


\begin{thebibliography}{99}
\bibitem{Notebom} It is worth noting that,
in order to trigger superradiant instabilities by bosonic fields in
spinning black-hole spacetimes, one must confine the extracted
rotational energy to a finite spacetime region in the vicinity of
the outer horizon. There are several known physical mechanisms that
can provide the confinement mechanism which is required in order to
support the exponentially growing superradiant bosonic fields: (1)
An artificial mirror placed inside the rotating black-hole
ergoregion \cite{PressTeu1,Ins1}, (2) A massive field whose bosonic
mass term acts as an effective confining mirror \cite{Ins2}, and (3)
A spinning black-hole spacetime with a non-asymptotically flat
reflecting boundary \cite{Ins3}.
\bibitem{Noteunr} It is important to stress the fact that, as originally proved in
\cite{Unr}, the wave equation which characterizes the dynamics of
acoustic sound modes in fluids systems [see Eq. (\ref{Eq4}) below]
is mathematically analogous to the well known Klein-Gordon equation
which characterizes the dynamics of massless scalar fields in curved
spacetimes.
\bibitem{Notepar} As explicitly proved in \cite{CCP},
the location of the inner radius $r_{\text{s}}$ depends on the
circulation of the fluid, its asymptotic mass density, and on the
physical parameters $k_{\text{p}}$ and $N_{\text{p}}$ [see Eq.
(\ref{Eq1})].
\bibitem{Notecir} The outer boundary $r=r_{\text{e}}$ of the acoustic ergoregion
is determined by the circle at which the circular velocity of the
fluid equals the propagation speed $c_{\text{s}}$ of acoustic sound
modes \cite{CCP}.
\bibitem{Notemm} Note that the decomposition (\ref{Eq6}) describes cylindrical perturbation
modes of the effective $(2+1)$-dimensional polytropic hydrodynamic
vortex, where the azimuthal index $m$ is an integer.
\bibitem{Notepa} Using the characteristic asymptotic behavior ${_2F_1}(a,b;c;z\to0)\to1$ of
the hypergeometric function \cite{Abram}, one finds that the radial
eigenfunction (\ref{Eq9}) is well behaved (bounded) at spatial
infinity: $\psi_m(x\to\infty;N_{\text{p}})\to A\cdot x^{1/2-m}$,
thus yielding the simple asymptotic behavior $\Psi(x\to\infty)\sim
x^{-m}$ [see Eq. (\ref{Eq6})] for the static linearized perturbation
modes of the polytropic hydrodynamic vortex.
\bibitem{Notenpm} Note that, in the physical regime $N_{\text{p}}>0$
that we consider in this paper, the dimensionless physical parameter
$N_{\text{c}}$ [see Eq. (\ref{Eq17})] is non-negative by definition.
\bibitem{Noteceil} Here $\left \lceil{x}\right \rceil$ is the ceiling
function of the variable $x$, which denotes the least integer that
is greater than or equal to $x$.
\bibitem{Notenp1} Note that $c-a-b\equiv 1-N_{\text{p}}$ in Eq. (\ref{Eq19}) [see the resonance
equation (\ref{Eq13})]. Thus, the requirement $\Re(c-a-b)>0$ in
(\ref{Eq19}) is satisfied for $N_{\text{p}}<1$.
\bibitem{Notenpb} Note that $c-b\equiv 1-{\bar N_{\text{c}}}$ in Eq. (\ref{Eq19})
[see the resonance condition (\ref{Eq13})].
\bibitem{Noteflor} Here $\left \lfloor{x}\right \rfloor$ is the
floor function of the variable $x$, which denotes the largest
integer that is smaller than or equal to $x$.
\bibitem{Notewwf} Note that, in the $0<N_{\text{p}}<1$ regime, we use the floor function
$\lfloor{{\bar N}_{\text{c}}}\rfloor$ because, as the value of the
dimensionless composed function ${\bar N_{\text{c}}}$ gradually
increases, a new resonant mode (which is characterized by the
dimensionless relation $x_{\text{c}}=1$) is added to the discrete
set of static modes only when ${\bar N_{\text{c}}}$ becomes a
positive integer [see Eqs. (\ref{Eq19}), (\ref{Eq20}), and
(\ref{Eq21})].
\bibitem{Notex1} Note that if
${{\sqrt{N^2_{\text{p}}+m^2(1+2N_{\text{p}})}+N_{\text{p}}-m}\over{2}}$
is an integer [that is, if ${\bar N}_{\text{c}}$ is an integer [see
Eq. (\ref{Eq21})]] with $0<N_{\text{p}}<1$, then one of the
$N_{\text{r}}$ acoustic resonant modes of the polytropic
hydrodynamic vortex [see the analytical formula (\ref{Eq22}) for the
number of marginally-stable resonances] is characterized by the
dimensionless supporting radius $x_{\text{c}}=1$.
\bibitem{Notenum} Note, in particular, that for a fixed value of the
azimuthal harmonic index $m$, one finds
$N_{\text{r}}=\lfloor{{{\sqrt{N^2_{\text{p}}+m^2(1+2N_{\text{p}})}+N_{\text{p}}-m}\over{2}}}\rfloor=0$
[see Eq. (\ref{Eq22})] in the $N_{\text{p}}\to 0^+$ limit of the
polytropic hydrodynamic vortex.
\bibitem{Notelst} Note that one finds from Eq. (\ref{Eq22}) the simple relations $N_{\text{r}}=0$ for
$N_{\text{p}}<N^{\text{min}}_{\text{p}}(m)$ and $N_{\text{r}}=1$ for
$N_{\text{p}}=N^{\text{min}}_{\text{p}}(m)$.
\bibitem{Notelt1} Note that, in agreement with the required inequality $N_{\text{p}}<1$
[see Eq. (\ref{Eq22})], one finds from (\ref{Eq23}) the relation
$N^{\text{min}}_{\text{p}}(m) <1$ for all integer values of the
azimuthal harmonic index $m$.
\bibitem{Notelcc} As emphasized above, static resonant modes of the polytropic hydrodynamic
vortex that satisfy the equality $2(m+1)/[m(m+1)+2]=N_{\text{p}}$
correspond to the limiting value $x_{\text{c}}=1$ of the supporting
cylinder radius.
\bibitem{Noteminn} Note that, according to the analytical formula (\ref{Eq22}),
the regime $N_{\text{p}}(m)<N^{\text{min}}_{\text{p}}(m)$ is
characterized by $N_{\text{r}}=0$.
\bibitem{Noteinf} Note, in particular, that the radial function
$\Phi(x)$ is characterized by the relations $\Phi(x=x_{\text{c}})=0$
and $\Phi(x\to\infty)\to0$ [see \cite{Notepa} and also Eqs.
(\ref{Eq11}) and (\ref{Eq27})].
\bibitem{Noteres} Note that the value of $\alpha^*(m)$ as given in
(\ref{Eq33}) respects the assumed inequalities (\ref{Eq27}) and
(\ref{Eq31}).
\bibitem{Noten0} Note that the case $n=0$ corresponds to the trivial relation $N_{\text{p}}(m)\equiv0$.
\bibitem{Noteok1} As a consistency check we note that the value $N_{\text{p}}=2/(m-2)$
respects the analytically derived lower bound (\ref{Eq23}).
\bibitem{Notem3} Note that the requirement $N_{\text{p}}>0$ implies
$m\geq3$ in this case.
\bibitem{Noteok2} It is worth noting that the value $N_{\text{p}}=4(m+2)/[m(m-1)-4]$
respects the analytically derived lower bound (\ref{Eq23}).
\end{thebibliography}
\end{document}